\documentclass[a4paper]{article}

\usepackage{amsfonts,amsmath,amssymb,bm,graphicx,hyperref,subfigure}
\usepackage{19lomcon}        % Proceedings volume layout metrics
\usepackage{cite}             % Smart range citations

\begin{document}

%%%%%%%%%%%%%%%%%%%%%%%%%%%%%%%%%%%%%%%%%%%%%%%%%%%%%%%%%%%%%%%%%%
% The preamble of the paper
%%%%%%%%%%%%%%%%%%%%%%%%%%%%%%%%%%%%%%%%%%%%%%%%%%%%%%%%%%%%%%%%%%

\title{RELATIVISTIC QUANTUM MECHANICS DESCRIPTION OF NEUTRINO SPIN-FLAVOR OSCILLATIONS IN VARIOUS EXTERNAL FIELDS}

\author{Maxim Dvornikov\email{maxdvo@izmiran.ru}}

\affiliation{Pushkov Institute of Terrestrial Magnetism, Ionosphere and Radiowave Propagation (IZMIRAN), 108840 Troitsk, Moscow, Russia}

\affiliation{Physics Faculty, National Research Tomsk State University, 36 Lenin Avenue, 634050 Tomsk, Russia}

% You may repeat \author and \affiliation as many times as necessary!

\date{}
% Print it out!
\maketitle

%%%%%%%%%%%%%%%%%%%%%%%%%%%%%%%%%%%%%%%%%%%%%%%%%%%%%%%%%%%%%%%%%%
% The preamble of the paper
%%%%%%%%%%%%%%%%%%%%%%%%%%%%%%%%%%%%%%%%%%%%%%%%%%%%%%%%%%%%%%%%%%

\begin{abstract}
We review the application of the relativistic quantum mechanics method for the description of neutrino oscillations for the studies of spin-flavor oscillations in background matter under the influence of a plane electromagnetic wave. Basing on the new exact solution of the Dirac-Pauli equation for a massive neutrino in the given external fields, we derive the transition probabilities for spin and spin-flavor oscillations. The obtained expressions are analyzed for different types of the neutrino magnetic moments. Our results are compared with findings of other authors.
\end{abstract}

The existence of the neutrino masses and the mixing between different neutrino types has been  confirmed by the experimental observation of neutrino oscillations.
% (see, e.g., Ref.~\cite{Ago18}).
The neutrino interactions with various external backgrounds, e.g., with matter, are known to affect the process of neutrino oscillations. Despite neutrinos are electrically neutral particles, they can have nonzero magnetic moments, which are of a pure anomalous origin. In this case, there is a possibility for neutrinos
to change both flavors and polarizations in an external electromagnetic
field. These transitions are called neutrino spin-flavor
oscillations. In the present work, we review our recent achievements, made in Refs.~\cite{Dvo18,Dvo19}, in the relativistic quantum mechanics description of neutrino spin-flavor oscillations in background matter under the influence of a plane electromagnetic wave.

Without loss of generality, we study the system of two flavor
neutrinos $(\nu_{\alpha},\nu_{\beta})$ with a nonzero mixing. For
example, we can take that $\nu_{\alpha}\equiv\nu_{\mu,\tau}$ and
$\nu_{\beta}\equiv\nu_{e}$. These neutrinos can electroweakly interact
with background matter consisting of electrons, protons, and neutrons.
The background matter is supposed to be nonmoving and unpolarized.
Moreover, we take that neutrinos have nonzero magnetic moments
and can interact with the external electromagnetic field $F_{\mu\nu}=(\mathbf{E},\mathbf{B})$.

The Lagrangian for the system of these neutrinos has the form,
\begin{equation}\label{eq:Lagrnu}
  \mathcal{L} =
  \sum_{\lambda\lambda'=\alpha,\beta}\bar{\nu}_{\lambda}
  \left[
    \delta_{\lambda\lambda'}\mathrm{i}\gamma^{\mu}\partial_{\mu} -
    m_{\lambda\lambda'} -
    \frac{M_{\lambda\lambda'}}{2}F_{\mu\nu}\sigma^{\mu\nu} -
    \frac{f_{\lambda\lambda'}}{2}\gamma^{0}(1-\gamma^{5})
  \right]
  \nu_{\lambda'},
\end{equation}
where $\gamma^{\mu}=\left(\gamma^{0},\bm{\gamma}\right)$, $\gamma^{5}=\mathrm{i}\gamma^{0}\gamma^{1}\gamma^{2}\gamma^{3}$,
and $\sigma_{\mu\nu}=\tfrac{\mathrm{i}}{2}[\gamma_{\mu},\gamma_{\nu}]_{-}$
are the Dirac matrices. The mass matrix $(m_{\lambda\lambda'})$ and
the matrix of magnetic moments $(M_{\lambda\lambda'})$ are independent
in general. The matrix of the effective potentials of the neutrino
interaction with matter is diagonal in the flavor basis: $f_{\lambda\lambda'}=f_{\lambda}\delta_{\lambda\lambda'}$.
The explicit form of $f_{\lambda}$ in the electroneutral matter can
be found in Ref.~\cite{DvoStu02}.
%as
%%
%\begin{equation}\label{eq:fnu}
%  f_{\nu_{e}} =
%  \sqrt{2}G_{\mathrm{F}}
%  \left(
%    n_{e}-\frac{1}{2}n_{n}
%  \right),
%  \quad 
%  f_{\nu_{\mu},\nu_{\tau}}=-\frac{1}{\sqrt{2}}G_{\mathrm{F}}n_{n},
%\end{equation}
%%
%where $G_{\mathrm{F}}=1.17\times10^{-5}\,\text{GeV}^{-2}$ is the
%Fermi constant and $n_{e,n}$ are the number densities of electrons
%and neutrons.

The nature of neutrinos can be revealed only if we transform the flavor
wave functions $\nu_{\lambda}$ to the mass eigenstates basis, $\nu = U \psi$, where $U$ is the mixing matrix, which depends on the vacuum mixing angle $\theta$. The neutrino mass eigenstates
$\psi_{a}$, $a=1,2$, having the masses $m_{a}$, are taken to be Dirac particles.
%In general
%situation, the matrices of magnetic moments $(\mu_{ab})=(U_{\lambda a})^{\dagger}(M_{\lambda\lambda'})(U_{\lambda'b})$
%and neutrino interaction with background matter $(V_{ab})=(U_{\lambda a})^{\dagger}(f_{\lambda\lambda'})(U_{\lambda'b})$
%are nondiagonal in the mass eigenstates basis.
%
%Using Eq.~(\ref{eq:nupsi}), we can rewrite the Lagrangian in Eq.~(\ref{eq:Lagrnu})
%in the following way:
%%
%\begin{equation}\label{eq:Lagrpsi}
%  \mathcal{L}=\sum_{ab=1,2}\bar{\psi}_{a}
%  \left[
%    \delta_{ab}
%    \left(
%      \mathrm{i}\gamma^{\mu}\partial_{\mu}-m_{a}
%    \right) -
%    \frac{\mu_{ab}}{2}F_{\mu\nu}\sigma^{\mu\nu} -
%    \frac{V_{ab}}{2}\gamma^{0}(1-\gamma^{5})
%  \right]
%  \psi_{b}.
%\end{equation}
%%
%One can see that the Dirac equations for different mass eigenstates,
%resulting from the Lagrangian in Eq.~(\ref{eq:Lagrpsi}), are coupled
%due to the presence of external fields.

The wave equations for $\psi_a$ have the form,
\begin{align}\label{eq:Direqab}
  \mathrm{i}\partial_{t}\psi_a = & \mathcal{H}_a \psi_a + \mathcal{V} \psi_b,
  \quad
  a,b = 1,2,
  \quad
  a \neq b,
  \notag
  \\
  \mathcal{H}_a & =
  -\mathrm{i}(\bm{\alpha}\nabla)+\beta m_a+
  \mu_{a}(\mathrm{i}\bm{\gamma}\mathbf{E} -
  \beta\bm{\Sigma}\mathbf{B})+\frac{V_{a}}{2}(1-\gamma^{5}),
  \notag
  \\
  \mathcal{V} & =
  \mu_{12}(\mathrm{i}\bm{\gamma}\mathbf{E} -
  \beta\bm{\Sigma}\mathbf{B})+\frac{V_{12}}{2}(1-\gamma^{5}),
\end{align}
where $\bm{\alpha}=\gamma^{0}\bm{\gamma}$, $\beta=\gamma^{0}$, and
$\bm{\Sigma}=\gamma^{5}\gamma^{0}\bm{\gamma}$ are the Dirac matrices, $(\mu_{ab})=(U_{\lambda a})^{\dagger}(M_{\lambda\lambda'})(U_{\lambda'b})$
and $(V_{ab})=(U_{\lambda a})^{\dagger}(f_{\lambda\lambda'})(U_{\lambda'b})$
are the matrices in the mass eigenstates basis.

The general solution of Eq.~\eqref{eq:Direqab} can be represented in the form~\cite{Dvo11},
\begin{align}\label{eq:Direqab}
  \psi_a(\mathbf{x},t) = \frac{1}{\sqrt{\varOmega}}
  \sum_n
  \left[
    a_{an}(t) u_{an}(\mathbf{x},t) + b_{an}(t) v_{an}(\mathbf{x},t)
  \right],
\end{align}
where $u_{an}$ and $v_{an}$ are the basis solutions of the diagonal part of Eq.~\eqref{eq:Direqab}, i.e., at the absence of the potential $\mathcal{V}$, depending on the set of quantum numbers $\{ n \}$, which, in its turn, is determined by the external fields, and $\varOmega$ is the normalization volume. In the relativistic quantum mechanics approach, used here, the $c$-number functions $a_{an}$ and $b_{an}$ are not the creation and annihilation operators~\cite{Dvo11}. Their form is chosen to account for the mixing potential $\mathcal{V}$ in Eq.~\eqref{eq:Direqab} and the initial conditions.

Now let us choose the particular configuration of the electromagnetic field. We suppose that the beam of neutrinos interacts with a plane electromagnetic wave propagating along the $z$-axis. First, we should find the exact solution of the diagonal part of Eq.~\eqref{eq:Direqab}. We use the Dirac matrices in the chiral representation. Omitting the index $a$ for brevity, one gets that this solution takes the form~\cite{Dvo19},
\begin{equation}\label{eq:psigensol}
  \psi=
  \exp
  \left(
    -\mathrm{i}\frac{V}{2}t+\mathrm{i}\mathbf{p}_{\perp}\mathbf{x}_{\perp}-
    \mathrm{i}\frac{\lambda}{2}u_{3} -
    \mathrm{i}\frac{\lambda}{2}\frac{p_{\perp}^{2}+m^{2}}{\lambda^{2}-V^{2}/4}u_{0}  
  \right)
  u,
\end{equation}
where
\begin{equation}\label{eq:basspinu}
  u^\mathrm{T} = \frac{1}{\sqrt{N}}
  \left(
    \frac{(p_{x}-\mathrm{i}p_{y})v_{1}-mv_{2}}{\lambda+V/2},
    v_{1},
    v_{2},
    -\frac{(p_{x}+\mathrm{i}p_{y})v_{2}+mv_{1}}{\lambda-V/2}
  \right),
\end{equation}
%\begin{equation}\label{eq:basspinu}
%  u = \frac{1}{\sqrt{N}}
%  \left(
%    \begin{array}{c}
%      \frac{(p_{x}-\mathrm{i}p_{y})v_{1}-mv_{2}}{\lambda+V/2}\\
%      v_{1}\\
%      v_{2}\\
%      -\frac{(p_{x}+\mathrm{i}p_{y})v_{2}+mv_{1}}{\lambda-V/2}
%    \end{array}
%  \right),
%\end{equation}
%
is the basis spinor. The normalization coefficient $N$ is given by
the condition $|u|^{2}=1$. In Eq.~\eqref{eq:psigensol}, we use the new variables $u_{0}=t-z$ and $u_{3}=t+z$, $\lambda$ is the quantum number~\cite{Dvo19}, and the symbol $\perp$ marks the vectors perpendicular to the wave propagation direction. The two component spinor $v=v(u_{0})=(v_{1},v_{2})^{\mathrm{T}}$ in Eq.~\eqref{eq:basspinu} obeys the equation, $\mathrm{i}\partial v/\partial u_0=(\bm{\sigma}^{*}\mathbf{R})v$, where $\bm{\sigma}$ are the Pauli matrices and
\begin{equation}\label{eq:R}
  \mathbf{R}=\mu\mathbf{B}+\frac{V}{2}\frac{m\mathbf{p}_{\perp}}{\lambda^{2}-
  V^{2}/4}-\frac{V}{4}
  \left(
    1+\frac{p_{\perp}^{2}-m^{2}}{\lambda^{2}-V^{2}/4}
  \right)
  \mathbf{e}_{z}.
\end{equation}
Here $\mathbf{e}_{z}$ is the unit vector along the wave propagation. In the following, we take that neutrinos propagate along the wave.

Let us consider the quasiclassical approximation. In this limit,
a neutrino moves along a trajectory, which is a straight line $z=\beta t$,
where $\beta=p_{z}/E$ is the neutrino velocity. Using this approximation, basing on Eqs.~\eqref{eq:psigensol} and~\eqref{eq:basspinu}, and considering ultrareletivistic neutrinos, we can reproduce the transition probabilities for spin oscillations obtained in Refs.~\cite{Dvo19,EgoLobStu00}.

Now we turn to the consideration of the neutrino spin-flavor oscillations $\nu_{\beta\mathrm{L}}\to\nu_{\alpha\mathrm{R}}$. First, we analyze the situation when $\mu_a \gg \mu$, where $\mu_a \equiv \mu_{aa}$ and $\mu \equiv \mu_{12}$. Considering the case when $V_{12}\ll |V_{aa}|$, one gets the analytical transition probability in the form~\cite{Dvo19},
\begin{equation}\label{eq:PLRbigmua}
  P_{\beta\mathrm{L}\to\alpha\mathrm{R}}(t)=\sin^{2}(2\theta)
  \left[
    \frac{1}{4}(A_{1}-A_{2})^{2}+A_{1}A_{2}\sin^{2}(\Phi t)
  \right],
\end{equation}
where
\begin{align}\label{eq:Aa}
  A_{a}(t) = & \frac{\mu_{a}B_{0}}{Z}(1-\bar{\beta})
  \sin(\varpi_a t),
  \nonumber
  \\
  \varpi_a & =
  \sqrt{\mu_{a}^{2}B_{0}^{2}(1-\bar{\beta})^{2}+
    \left[
      \frac{V_{aa}}{2}+\frac{g\omega}{2}(1-\bar{\beta})
    \right]^{2}},
\end{align}
is the amplitude of spin oscillations within one mass eigenstate,
$\Phi=\Phi_{\mathrm{vac}}+(V_{11}-V_{22})/2$ is the phase of neutrino
flavor oscillations accounting for the matter contribution, $\Phi_{\mathrm{vac}}=\delta m^{2}/4p_{z}$
is the phase of neutrino oscillations in vacuum, $\delta m^{2}=m_{1}^{2}-m_{2}^{2}$, and $\bar{\beta}$
is the center of inertia velocity of two mass eigenstates.

\begin{figure}
  \centering
  \subfigure[]
  {\label{2a}
  \includegraphics[scale=.28]{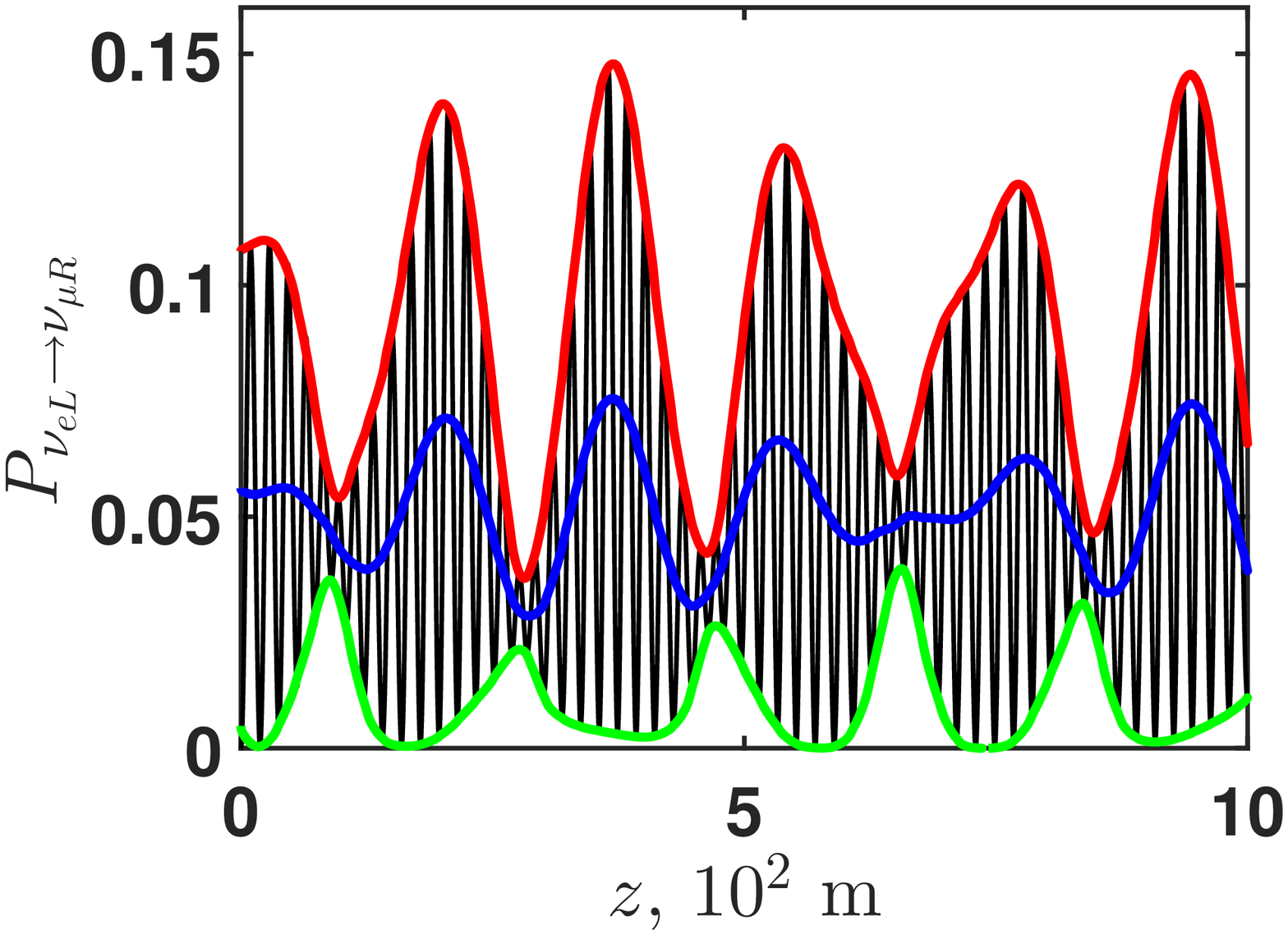}}
  \hskip-.6cm
  \subfigure[]
  {\label{2b}
  \includegraphics[scale=.28]{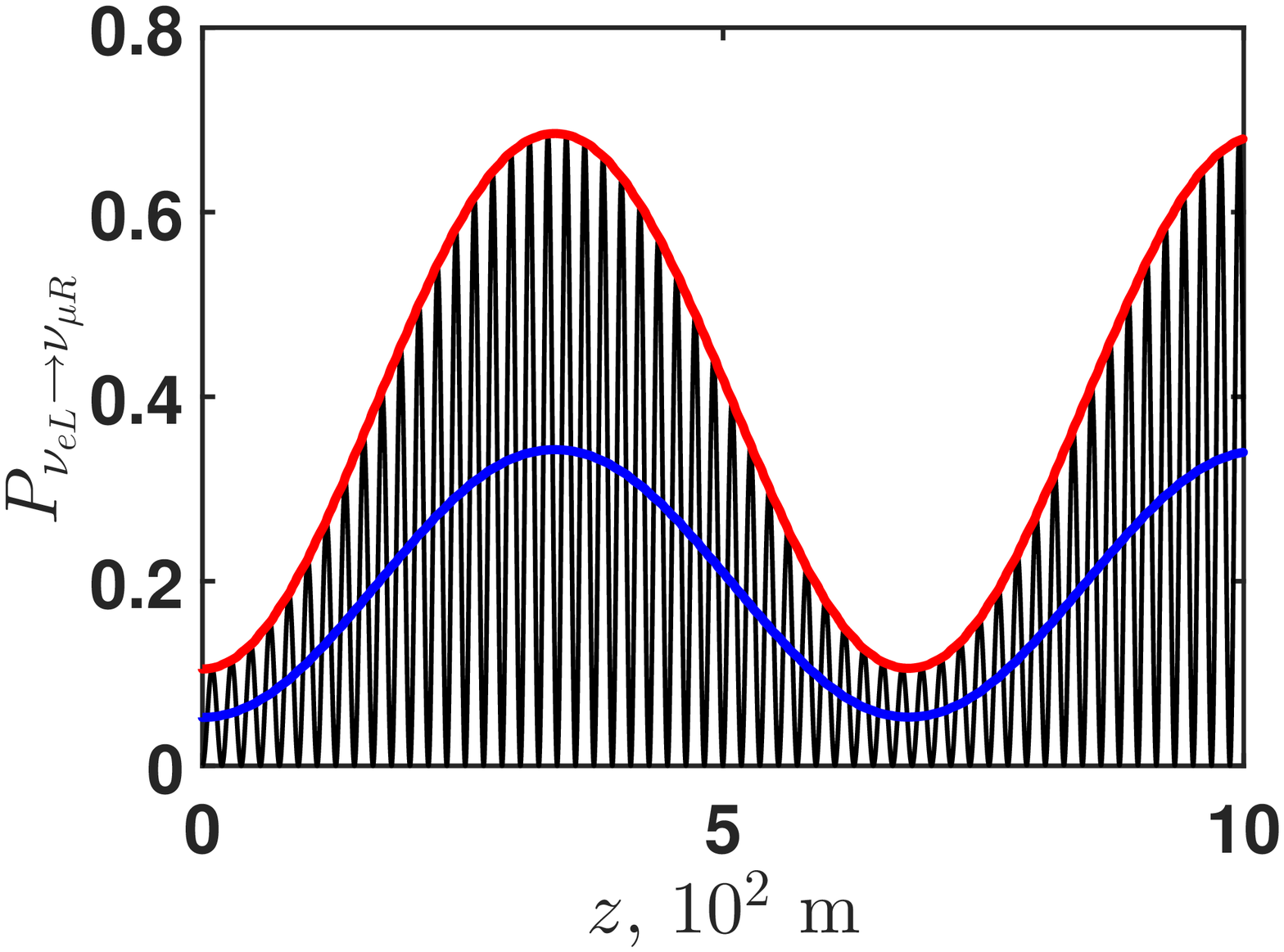}}
  \protect
  \caption{The transition probabilities for $\nu_{e\mathrm{L}}\to\nu_{\mu\mathrm{R}}$
  oscillations in the electroneutral hydrogen plasma when particles
  interact with the electromagnetic wave having $B_{0}=10^{18}\,\text{G}$
  and $\omega=10^{13}\,\text{s}^{-1}$ versus the distance $z=\bar{\beta}t$
  passed by the neutrino beam. The parameters of neutrinos are
  $\delta m^{2}=7.59\times10^{-5}\,\text{eV}^{2}$,
  $\theta=0.6$, $p_z=1\,\text{keV}$, $m_a \sim 1\,\text{eV}$,
  and $\mu=10^{-11}\mu_{\mathrm{B}}$.
  (a) $n_{e}=10^{29}\,\text{cm}^{-3}$; and 
  (b) $n_{e}=10^{27}\,\text{cm}^{-3}$. Red and blue lines are the upper
  envelope functions and the averaged transition probabilities. The
  green line in panel~(a) is the lower envelope function.\label{fig:eLmuR}}
\end{figure}

The transition probabilities in the situation when $\mu \gg \mu_a$ can be obtained only numerically for the arbitrary $\theta$ and $(V_{ab})$. The effective Hamiltonian, which drives spin-flavor oscillations in this case, has been derived in Ref.~\cite{Dvo19}.
Using that result, in Fig.~\ref{fig:eLmuR}, we show the transition probabilities $P_{\nu_{e\mathrm{L}}\to\nu_{\mu\mathrm{R}}}(z)$,
the upper and lower envelope functions, and the averaged transition
probability.

One can see in Fig.~\ref{2a},
that the averaged transition probability oscillates near $5\%$ value. In Fig.~\ref{2b}, we depict $P_{\nu_{e\mathrm{L}}\to\nu_{\mu\mathrm{R}}}(z)$
for lower matter density $n_{e}=10^{27}\,\text{cm}^{-3}$, which is
very close to the value expected in the inner part of an accretion disk in some models of gamma ray bursts. The transition probability
in this case reproduces the result in Ref.~\cite{Dvo18}, where spin-flavor
oscillations $\nu_{e\mathrm{L}}\to\nu_{\mu\mathrm{R}}$ were studied
at the absence of the matter contribution. Comparing Figs.~\ref{2a}
and~\ref{2b}, one can see that the lower matter density
is, the higher transition probability is. Thus, one does not expect
the appearance of a resonance in neutrino spin-flavor oscillations
in matter under the influence of a plane electromagnetic wave, as
claimed in Ref.~\cite{EgoLobStu00}. The highest transition probability
can be observed when neutrinos do not interact with background matter.

\section*{Acknowledgments}

%\paragraph{Acknowledgments}
I am thankful to the Russian Science Foundation for the support (Grant No.~19-12-00042).

%%%%%%%%%%%%%%%%%%%%%%%%%%%%%%%%%%%%%%%%%%%%%%%%%%%%%%%%%%%%%%%%%%
% References
%%%%%%%%%%%%%%%%%%%%%%%%%%%%%%%%%%%%%%%%%%%%%%%%%%%%%%%%%%%%%%%%%%

\end{document}